\documentclass[11pt,a4paper]{article}

\usepackage{amsmath}
\usepackage{amsfonts}
\usepackage{amssymb}
\usepackage{graphicx}
\usepackage[utf8]{inputenc}
\usepackage[T1]{fontenc}
\usepackage{mathrsfs,amsmath} 
\usepackage{subcaption}
\usepackage[labelfont=bf]{caption}
\usepackage[a4paper]{geometry}
\usepackage{authblk}
\usepackage{booktabs}
\usepackage[framemethod=TikZ]{mdframed}
\usepackage{lipsum} 
\usepackage{threeparttable}
\usepackage{float}
\usepackage{multirow}
\usepackage{rotating}
\usepackage[numbers,sort&compress]{natbib}
\usepackage{hyperref}
\usepackage{longtable}
\usepackage{lineno}

\usepackage{color}
\definecolor{myorange}{RGB}{240, 96, 0}

\definecolor{myblue}{RGB}{30,144,255}

\definecolor{mypurple}{RGB}{128, 0, 128}

\definecolor{mygreen}{RGB}{0, 97, 62}

\title{Using a Cognitive Network Model of Moral and Social Beliefs to Explain Belief Change\\[0.4em]\large Short title: Network Model of Moral and Social beliefs}

\author[1,*,$\dag$]{Jonas Dalege }
\author[1,$\dag$]{Tamara van der Does}
\affil[1]{Santa Fe Institute, 1399 Hyde Park Rd, Santa Fe, NM 87501.}
\affil[*]{corresponding author: j.dalege@santafe.edu}
\affil[$\dag$]{Equal authorship, order determined by universe splitter.}
\date{}

\begin{document}
\maketitle

\begin{abstract}
Scepticism towards childhood vaccines and genetically modified food has grown despite scientific evidence of their safety. Beliefs about scientific issues are difficult to change because they are entrenched within many related moral concerns and beliefs about what others think. We propose a cognitive network model which estimates the relationships, dissonance, and randomness between all related beliefs to derive predictions of the circumstances under which beliefs change. Using a probabilistic nationally representative longitudinal study, we found support for our model's predictions: Randomness of the belief networks decreased over time, for many participants their estimated dissonance related positively to their self-reported dissonance, and individuals who had high estimated dissonance of their belief network were more likely to change their beliefs to reduce this dissonance. This study is the first to combine a unifying predictive model with an experimental intervention and sheds light on dynamics of dissonance reduction leading to belief change.\end{abstract}

\textbf{Teaser:} Using a longitudinal study, we find that high estimated dissonance between beliefs is associated with belief change.

\pagebreak


\section*{Introduction}
The World Health Organization (WHO) lists vaccination hesitancy as one of the ten greatest threats to global health \cite{WHO2019vac}. Erroneous beliefs regarding vaccines, which are somewhat common in the US \cite{Funk2017}, can accelerate or even re-ignite the spread of diseases globally. In another recent report, the WHO also points out that around 45\% of deaths among children under 5 years of age are linked to undernutrition \cite{WHO2020malnu}. Even though the scientific community has confirmed that currently approved Genetically Modified (GM) crops are safe and could provide higher yields \cite{NAS2016GM}, many US Americans are sceptical about this technology \cite{Funk2015, Funk2016a}. Other beliefs inconsistent with the scientific consensus, such as climate change denial, can have similar detrimental consequences for society. We need to understand how these sceptical beliefs about scientific issues can be changed in order to develop successful public science communication. 

In this paper, we use a cognitive network model inspired by statistical physics to understand change in beliefs about GM food and childhood vaccines. We consider attitudes towards GM food and childhood vaccines as networks of connected beliefs \cite{Dalege2016, Monroe2008, VanOverwalle2005} and use this model to precisely estimate the relationships, dissonance, and randomness between all beliefs. Using data data from a longitudinal nationally representative study with an educational intervention, we test if predictions derived from our cognitive network model can explain under which circumstances individuals are more likely to change their beliefs over time and shed light on the dynamic nature of dissonance reduction leading to belief change. By combining a unifying predictive model with an experimental longitudinal dataset, we expand upon the strengths of earlier investigations into science communication and belief change dynamics, as we describe in the next paragraphs.

Previous applied research on beliefs about GM food and childhood vaccines has found that scepticism about their safety is shaped both by related moral beliefs (e.g., care for others, concerns about the environment, importance of naturalness, and purity) and perceived beliefs of trusted social groups, such as doctors or family members \cite{Amin2017, Clifford2016, Kahan2011, Miton2015, Rutjens2018, Scott2016a, Johnson2020}. Studies focusing on changing these beliefs have therefore tried to vary the framing of the factual information and the source of information, with mixed success \cite{Betsch2011, Horne2015, Nyhan2014, Rothman1997}. This literature sheds light on important related beliefs (both moral and social) as determinants of beliefs about GM food and childhood vaccines. However,  these empirical studies tend to focus on specific interventions and populations \cite{Dube2015} and do not draw on a unifying model to understand the processes underlying belief change more generally.

Mirroring findings from applied research on GM food and childhood vaccines, general models of belief change have identified two important sets of factors as consistently important for belief change (for a review, see \cite{Galesic2020}). First, individuals hold many related personal beliefs, such as moral beliefs. In social psychology, the concept of dissonance was developed to understand when and why individuals might change their beliefs when they are incoherent with each other \cite{Elliot1994, Festinger1962}. Within this approach, incoherent beliefs translate into feelings of dissonance and belief change if attention is paid to these beliefs. Building on this concept of dissonance, more recent research has modelled the relationship between personal beliefs using cognitive network models to predict belief dynamics \cite{Dalege2019, Dalege2016, Kunda1996, Monroe2008, VanOverwalle2005}. Second, individuals' beliefs are shaped by their social networks. In statistical physics, models of opinion dynamics can predict change over time within a social network \cite{Redner2019, Castellano2009}. These models, however, generally do not take into account that the beliefs held by one's social network do not directly influence one's own beliefs. Instead, their influence is mediated by how one perceives beliefs in one's social network \cite{Cialdini1998, Festinger1954, Ajzen1977, Helbing1995b}, which implies that perceptions of beliefs in one's social network provide information above and beyond actual beliefs in one's social network \cite{galesic2021human}. For example, a person might overestimate how liberal their friends are, and thus become more liberal themselves, just because their liberal friends voice their political position more firmly than their moderate friends. 

In recent years, a few belief change models were developed to focus specifically on the interaction between related personal beliefs (e.g., moral beliefs) and beliefs about one's social network (social beliefs). In general, these models have either focused on dissonance between all moral and social beliefs using the statistical physics concept of energy \cite{Galesic2019}, or on network imbalance between personal and social beliefs (e.g., belief A is positively connected to beliefs B and C, but belief B and C are negatively connected) \cite{Rodriguez2016, Schweighofer2020, Li2019}. Models based on dissonance were able to predict belief change using estimated energies from reported moral and social beliefs \cite{vanderDoes2021}. However, these models did not take the network structure of related moral beliefs into account. Models based on network imbalance were able to predict distributions of beliefs \cite{Schweighofer2020} and provide an explanation of how minorities can convince majorities \cite{Rodriguez2016} but empirical tests of whether these model can predict belief change are still lacking. In the following sections, we present our cognitive network model before moving to the results section.  

Our cognitive network model integrates both moral and social beliefs to empirically predict belief change dynamics. We extend the recent Attitudinal Entropy (AE) framework \cite{Dalege2018}, a model inspired by statistical physics which conceptualises individual overall attitudes as networks comprising of different beliefs or ``spins'' (i.e., elements of the network that can take different values) that are of binary nature. Within this framework, individuals change their beliefs towards more consistency, so as to reduce their attitudinal entropy (a measure of unpredictability of the whole belief network). We build upon this model to include both moral and social beliefs and by generalising to beliefs that can take any value between -1 (complete disagreement) and 1 (complete agreement). We discuss the implications of having continuous beliefs in the results section. In the next three paragraphs, we explain how each major concept in our model relates to psychological constructs and belief change before discussing our empirical predictions. 

The main concepts of our cognitive network model, their proposed psychological meaning, and the way we estimate them are listed in Table \ref{tab:1}. Couplings between both moral and social beliefs represent the strength and sign (positive or negative) of the relationships between beliefs. These couplings are estimated using partial correlations between nodes in the network. The strength of the couplings within the network determines the probability of beliefs being in consistent states. A network with strong couplings is more likely to have consistent beliefs, while a network with weak couplings is more likely to have inconsistent beliefs. The sign of the couplings in the network determines which belief states (spins) can be regarded as consistent. Let's assume, for example, that the belief that vaccines are safe and the belief that they are effective are positively connected, and that the belief that vaccines are safe is negatively connected to the belief that pharmaceutical companies are only interested in making money regardless of patients' health. The belief network would be highly consistent if the individual agrees with the former two beliefs and disagrees with the later belief (or, conversely, disagrees with the former two beliefs and agrees with the later belief). 

\begin{table}
\centering
\caption{Overview of cognitive network model parameters within the statistical physics framework, and their corresponding psychological constructs and methods of estimation.}
\label{tab:1}
\begin{tabular}{p{0.20\linewidth} | p{0.35\linewidth} | p{0.45\linewidth}} \hline
\bf{Statistical physics term}  & \bf{Psychological construct}  & \bf{Estimation}  \\ 
\hline
Coupling $\omega_{ij}$ 
& Relationship between two beliefs. 
& Partial correlation between $b_i$ and $b_j$ controlled for all other beliefs. \\
\hline
Energy $H$ 
& Dissonance.
& Measures inconsistency (high energy) or consistency (low energy) of beliefs given estimated network structure. Sum of weighted products of self-reported belief scores, $-\omega_{ij} b_i b_j$.  \\
\hline
Temperature $1/\beta$ 
& Subsumes several processes that increase randomness and disorder of belief networks such as lack of attention and no thought directed to the belief network.
& Measures inverse of interdependence between beliefs. Average of the diagonal of the inverse covariance matrix of beliefs. In other words, average of belief-specific scaling values of $b_i$ and $b_j$ which are estimated in order to transform $\omega_{ij}$  into measured correlations.  \\
\hline
\end{tabular}
   \end{table}

Energy can be understood as a formalisation of the psychological concept of dissonance. Dissonance refers to the actual inconsistency between beliefs and is translated into felt dissonance if enough attention is given to these beliefs. According to classic psychological theories and the AE framework, felt dissonance leads to belief change because individuals want their beliefs to be in a consistent state \cite{Festinger1962, Dalege2018}. Energy is measured as the sum of the products of each pair of beliefs (spins) and their relationship (coupling) multiplied by -1. A consistent network has low energy and an inconsistent network has high energy. 

Temperature can be interpreted using several psychological processes that increase randomness and disorder, such as lack of attention and no thought directed to beliefs, with lower temperature corresponding to higher attention and thought. Temperature is estimated using the average of scaling values transforming individual couplings (estimated partial correlations) into measured correlations. Therefore, temperature relates to the interdependence between beliefs: High measured correlations between beliefs result in low estimated temperature, while low measured correlations between beliefs result in high estimated temperature. Temperature influences belief change through its scaling of the couplings. Lower temperatures increase the impact of couplings on the belief states, while higher temperatures reduce the impact of the couplings. 
   
Relationships between different concepts in our cognitive network model can be expressed by Equations \ref{eq:1} and \ref{eq:2}:		
\begin{equation} \label{eq:1}
H_i = - \sum_{j:j\ne i} \omega_{ij} b_{i} b_{j}
\end{equation}

where $H_i$ is the energy (dissonance) of a belief, $b_i$ is the value of belief $i$ in the belief network (moral or social), and $\omega_{ij}$ is the coupling (relationship between beliefs) between $b_i$ and another belief in the network, $b_j$. The energy of one belief $H_i$ is the sum of this beliefs' individually weighted product with all other beliefs in the network. In this paper we focus on the couplings, thus omitting the external field from the energy equation. We discuss this and other assumptions in detail in the result section. The conditional probability that a given belief will change its state from its current state to a different state is 
\begin{equation} \label{eq:2}
P(b_{i}\rightarrow b_{i}')= 1/(1 +e^{\beta \Delta H_i})
\end{equation}
where   $\Delta H_i  =  H_i' - H_i$ is the change in energy between the two belief states ($H_i'=- \sum_{j:j\ne i} \omega_{ij} b_{i}' b_{j}$) and $\beta$ is the inverse temperature of the network. The probability of belief changing state from $b_i$ to $b_i'$ increases with (a) the difference in energy between the new state and the current state ($\Delta H_i$)  and (b) the reduction in temperature of the whole network ($\beta$). The lower the temperature, the higher the probability of belief change towards a new state with lower energy. 

\begin{figure}
    \centering
\includegraphics[width=\textwidth]{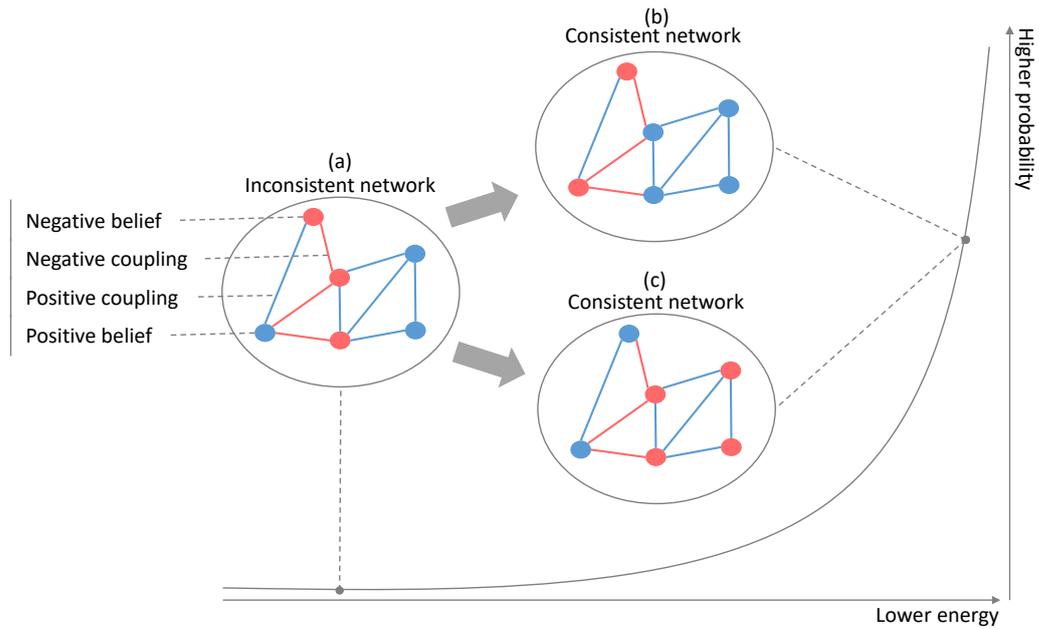}
\caption{\textbf{Cognitive network model of belief change.} A belief network will change over time to achieve higher consistency. Network (a) is inconsistent because it has similar beliefs connected by negative couplings and opposite beliefs connected by positive couplings. The inconsistent network (a) will change to achieve higher consistency, either towards (b) or (c). These changes reflect the higher probability in equilibrium of a low energy network compared to a high energy network, given a temperature less than infinity.}
\label{fig:energy}
\end{figure}

The whole belief network has energy $H=\sum_{i}H_i$, summing the energies of all the beliefs in the network. We expect beliefs to change towards more consistency of the whole network (Figure \ref{fig:energy}). This reflects the expected high probability of consistent networks (low energy states) in equilibrium. As shown in Figure \ref{fig:energy}, there are multiple ways for beliefs to change in order to achieve consistency. To note, our cognitive network model proposes that beliefs are more likely to move to a consistent state if the network estimated temperature is less than infinity (i.e., when at least some attention is directed at the beliefs). As temperature increases, beliefs are less likely to move to a consistent state, until there is no difference in probability between consistent and inconsistent states when temperature is infinitely high.

Three empirical predictions can be drawn from our cognitive network model. First, in our cognitive network model, temperature measures the inverse of interdependence between beliefs (Table \ref{tab:1}) and therefore should relate to processes that decrease disorder and randomness, such as attention to one's beliefs. This leads to the prediction that attention to beliefs should lead to lower estimated network temperatures. Second, our model holds that belief network energies are a formalisation of dissonance. Individuals with high energy should thus also have high feelings of dissonance. We therefore predict that estimated energies should be positively related to self-reported feelings of dissonance. Finally and most crucially, belief change is predicted to be more likely when individuals try to achieve higher consistency between their beliefs in order to reduce these feelings of dissonance. Our third prediction is therefore that individual energies should predict belief change and belief change should be associated with a process of lowering energies. In order to test these three predictions, we need longitudinal data on beliefs over time from which we can estimate an empirical model of belief networks drawn from our cognitive network model. Investigating the dynamics of dissonance reduction goes above and beyond the usual investigations into cognitive dissonance, which typically only investigate the consequences of inducing dissonance \cite{aronson1959effect, egan2007origins, harmon2002testing}. Here, we investigate how dissonance interacts with receiving new information on a topic and whether such new information leads to reconfiguration of one's beliefs leading to lower dissonance.



\section*{Results}

We used a nationally representative longitudinal study of beliefs about GM food and childhood vaccines to test implications of our cognitive network model (see Methods for details on the study design and questionnaire). This study included questions about both moral beliefs related to the safety of each technology (e.g., GM food [Childhood vaccines] are beneficial to children, GM food [Childhood vaccines] are part of our tradition) and social beliefs about their safety (e.g., \% of medical doctors believe GM food [Childhood vaccines] is [are] safe, \% of my family and close friends believe GM food/Childhood vaccines is [are] safe). We assessed these beliefs four times across three waves of data collection (over an average of 30 days): once in the first and third wave and twice in the second wave (before and after the intervention). In the second wave, we presented individuals with an educational intervention about the safety of GM food and vaccines, quoting reports from the National Academies of Sciences. A total of 979 individuals participated in all three waves and answered all relevant questions for the study.

\begin{figure}
    \centering
\includegraphics{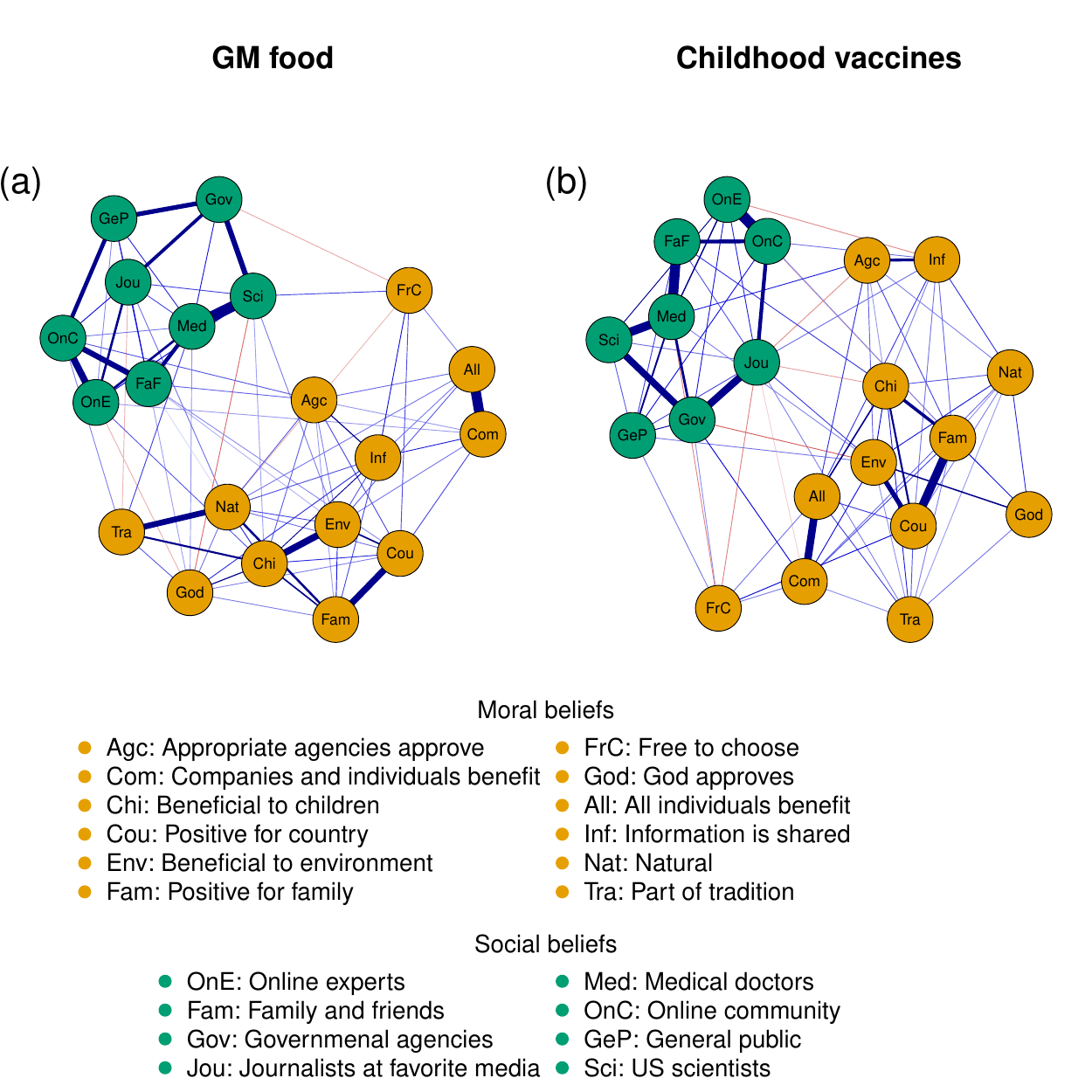}
\caption{\textbf{Belief networks for GM food (a) and childhood vaccines (b)}. Beliefs include moral (orange nodes) and social beliefs (green nodes). Edges represent couplings (partial correlations) between two beliefs controlled for all other beliefs. Blue (red) edges represent positive (negative) couplings and the widths of the edges correspond to the strength of the couplings. The strength of the couplings ranged from 0.02 (between the beliefs "Chi" and "Fam") to 0.30 (between the beliefs "Med" and "Sci") for GM food and from 0.02 (between the beliefs "Com" and "Jou") to 0.28 (between the beliefs "OnE" and "OnC"), N=979.}
\label{fig:networks}
\end{figure}

\subsection*{Estimation of the cognitive network}

 To fit our cognitive network model to empirical data, we focused on changes in all moral and social beliefs after removing variations explained by individual-level and time-level differences (see Methods for details). We also made four key assumptions. First, we assumed that Gaussian distributions are appropriate for our data (see empirical support in Supplementary Materials Figure \ref{fig:normal}), which allowed us to estimate a Gaussian Graphical Model (GGM). Second, our data represents equilibrium distributions. While individual beliefs can change, we assumed a fixed distribution of all the beliefs in the  belief network. Third, in contrast to what is typically assumed for GGM models, we assumed that individuals are not motivated to move to the global mean, because external fields vary between individuals. We think that this assumption is appropriate, because individuals have different dispositions for their beliefs, which makes it likely that they are motivated to move to different belief states. Fourth, we assumed that couplings in the estimated group-level belief networks are representative for the couplings at the individual level (see empirical support in Supplementary Materials Figure \ref{fig:indivpop})

Our model can be specified in several different ways when fitted to empirical data. We did not make any assumptions on (a) whether the couplings, the external fields, and/or temperature vary over time and (b) whether the networks are sparsely or densely connected. To investigate if our constructs vary over time and whether temperature drops during the time course of the study, we fitted different specifications of our model on the four time points. These specifications focused on constraining partial correlations to be equal across time points, external fields (mean values) to be equal across time points, and temperature to be equal across time points (see Methods for details on network estimation). Additionally, we tested whether the data can be captured best by a dense network (all beliefs are directly connected to all other beliefs) or a sparse network (some beliefs are not directly connected). The results indicated that a sparse network with equal partial correlations and equal external fields between time points and varying temperature across time points fitted the data best. This implies that the network structure remained constant throughout time, but that the interdependence between beliefs varied.  

The estimated group-level networks for beliefs regarding GM food and childhood vaccines are shown in Figure \ref{fig:networks}a and \ref{fig:networks}b. In both networks the moral and social beliefs were connected to each other but formed two distinct clusters. Most beliefs were positively connected but there were some negative connections as well. For example, there is a negative estimated coupling between the belief that scientists think GM food is safe and the one that God approves of GM food. The GM food network was more densely connected than the childhood vaccines network, indicating that beliefs toward childhood vaccines were more independent from each other. This might indicate that individuals formed more nuanced beliefs toward childhood vaccines than toward GM food. In addition to estimating couplings between beliefs, we also estimated the overall temperature of the networks over time.

\subsection*{Decrease in temperature over time}

According to our cognitive network model, attention to beliefs should lead to lower temperature. We expected that temperature of the network would decrease as individuals took part in the longitudinal survey about their beliefs related to the safety of GM food and childhood vaccination, because filling in these questionnaires should lead to a higher amount of attention directed at these beliefs. This higher amount of attention is expected to result in lower temperature and therefore higher interdependence between beliefs. In addition, an estimated network temperature that is not infinitely high, or with some interdependence between beliefs, is necessary for the relationship between energy and belief change to hold. Systems with temperature lower than infinity are more likely to move from inconsistent states to consistent states, while networks in infinitely high temperature have the same probabilities of changing to any state. 

\begin{figure}
    \centering
\includegraphics[scale = .8]{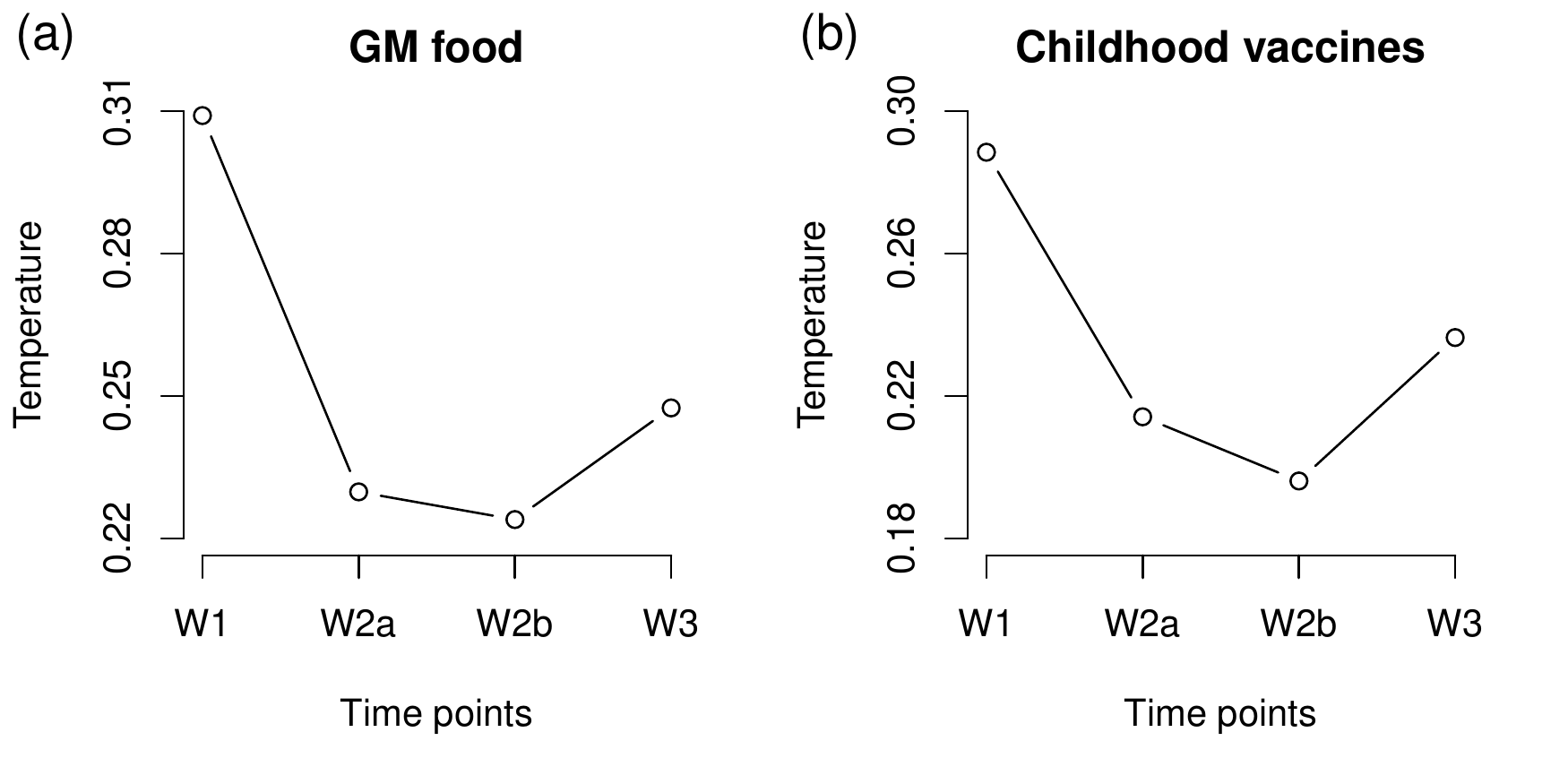}
\caption{\textbf{Changes in estimated temperature of belief networks through time}. Estimated temperature over time for GM food (a) and childhood vaccines (b), N=979.}
\label{fig:temperature}
\end{figure}

As can be seen in Figure \ref{fig:temperature}a and \ref{fig:temperature}b, the temperatures of both belief networks decreased with time, implying that all beliefs became more interdependent during our GM food and childhood vaccines studies. This confirms our first prediction that attention to beliefs would lead to a decrease in temperature. The sharpest decrease in the temperature was observed between the first and second time point, implying that the sharpest increase in the interdependence between beliefs was observed between the first and second measurement. It is noteworthy that simply administering a questionnaire has the strongest impact on temperature. A relatively low temperature between wave 2a (beliefs measured before the intervention) and wave 2b (beliefs measured after the intervention) means beliefs are  more likely to move from high energy states to low energy states. In psychological terms, more attention directed at the beliefs during the intervention resulted in a closer correspondence between dissonance and felt dissonance. This, in turn, should lead to belief change. Next, we test whether estimated energies indeed relate to self-reported dissonance. 

\subsection*{Correlation between estimated energy and self-reported dissonance}

\begin{figure}
    \centering
    \includegraphics[scale=0.7]{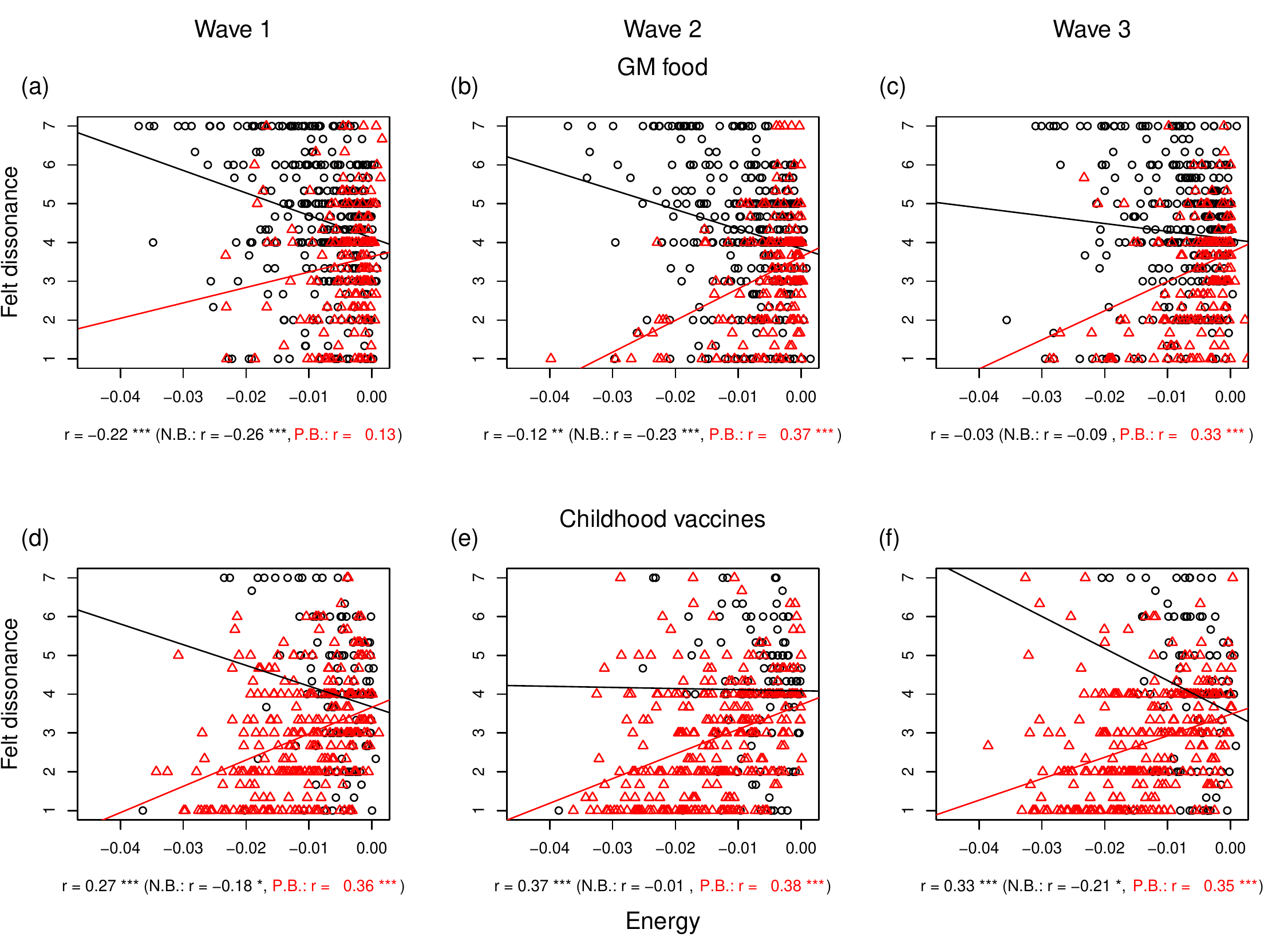}
    \caption{\textbf{Relationship between belief network energies and self-reported felt dissonance.} Black dots represent individuals who had belief sum scores lower than 0, indicating negative beliefs. Red triangles represent individuals who had belief sum scores equal or higher to 0, indicating neutral or positive beliefs. N.B.: Correlation estimates for individuals holding negative beliefs. P.B.: Correlation estimates for individuals holding neutral or positive beliefs, N=979.}
   \label{fig:scatterplots}
\end{figure}

According to our model, individual belief network energies should be positively related to self-reported feelings of dissonance. We calculated individual energies by summing each individual's actual beliefs weighted by the estimated couplings (See Methods for more details on their calculations). We measured self-reported felt dissonance in each of the three waves (but only one time in wave 2, after the educational intervention) and correlated them with the energy based on the beliefs at the same time point. We expected a positive relation between energy and self-reported dissonance. Indeed, with some attention directed to the subject at hand, individuals should be  sensitive to beliefs that might be contradictory from one another and report feelings of being uncomfortable, uneasy, and bothered \cite{Festinger1954, Elliot1994}.

The prediction that belief network energies and self-reported dissonance correlate positively received support among some groups of participants. The relationship between energies and self-reported dissonance in each wave, separating participants who first held positive or negative beliefs about GM food or childhood vaccines, is shown in Figure \ref{fig:scatterplots}a-f. Regarding beliefs toward GM food, considering all participants together, the correlations between belief network energies and dissonance did not follow a clear pattern and were mostly of weak magnitude. However, when considering participants with negative beliefs and those with positive beliefs separately, we found some interesting trends. There was a positive relationship between belief network energies and self-reported dissonance for participants holding somewhat positive beliefs about either topic. The relationship did not hold for participants with negative beliefs. Regarding beliefs toward childhood vaccines, considering all participants together, the correlations between belief network energies and dissonance were positive and of moderate magnitude. However, similar to the beliefs toward GM food, this relation was only found for individuals with positive beliefs. 
 
We believe that the lack of relationship between estimated energies and self-reported dissonance for participants who generally hold negative views towards GM food or vaccines is due to our measurement of felt dissonance, which might have been to unspecific, and its relationship with overall beliefs. Indeed, felt dissonance was strongly influenced by one's original beliefs. Across studies and waves (Figure \ref{fig:scatterplots}a-f), holding negative views towards vaccines was associated with higher felt dissonance. Participants holding negative beliefs toward GM food and childhood vaccines probably felt dissonance due to their impression that they were taking part in a study run by individuals with different beliefs. Participants were aware that the study was run by scientists, who are likely to hold positive beliefs toward GM food and childhood vaccines. Therefore, these participants probably experienced dissonance due to their beliefs being inconsistent with their impression of who created the questionnaire, and not due to the inconsistency of their own beliefs. While our model was estimated removing overall individual differences in beliefs, self-reported felt dissonance was inconsistent between people with different beliefs about GM food and vaccines. This also explains some differences between the GM food and vaccines study. We only observed a positive correlation between dissonance and energies for all participants in the childhood vaccines group, probably because there were more participants holding negative beliefs about GM food than about childhood vaccines. 

To summarise, our prediction that belief network energies relate positively to self-reported dissonance received mixed support. While this relation was found for individuals with positive beliefs regarding GM food and childhood vaccines, it did not hold for individuals with negative beliefs. It is likely that this was caused by the specific phrasing of the dissonance questions.

\subsection*{Energies predicting belief change}

Finally, our model predicts that individual energies should predict belief change and belief change should lead to lower energies. Most of our participants changed their beliefs over time, but not always in the expected direction. We measured belief change in moral beliefs by comparing the mean of all moral beliefs pre- and post-experimental intervention. We focused on moral beliefs because they reflect participants' own opinions and attitudes towards GM food and childhood vaccines to a larger extent than their social beliefs. In our sample, 46\% on average had changes in their networks towards more accepting (positive) beliefs regarding GM food and childhood vaccines. Even though the education intervention showed evidence for the safety of GM food and childhood vaccines, 42\% of our participants changed their beliefs on average towards more scepticism (negative beliefs). This type of backlash is quite common in studies of beliefs about GM food and childhood vaccines \cite{Fernbach2019, Nyhan2014}. Beliefs regarding GM food were more likely to change negatively compared to beliefs regarding childhood vaccines. However, according to our cognitive network model (Figure \ref{fig:energy}), the relationship between network energies and belief change should hold regardless of the direction of belief change.

\begin{figure}
    \centering
\includegraphics[scale=0.65]{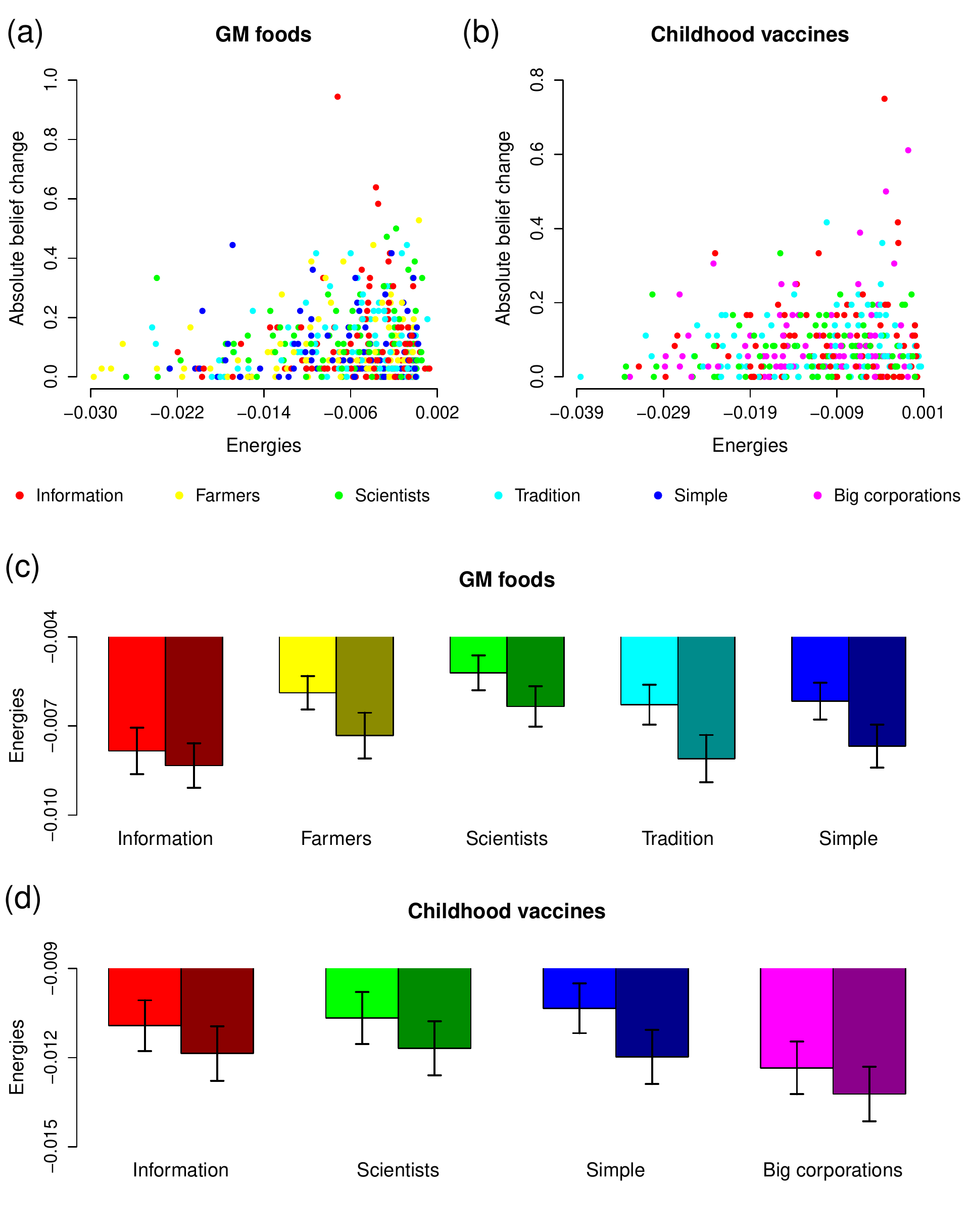}
\caption{\textbf{Relationships between belief network energies before interventions and absolute change of beliefs during interventions.} (a) and (b) show the relation between energies and absolute belief change for GM foods, and childhood vaccines, respectively, where each dot represents a participant. (c) and (d) show the belief network energies before (lighter saturation of bar colours) and after (darker saturation of bar colours) the interventions. (a) and (b) show that belief network energies correlate with whether individuals will change their beliefs during interventions. (c) and (d) show that these changes in beliefs are driven towards lower energy states. Colours of points and bars correspond to interventions. Error bars in (c) and (d) indicate means $\pm$ 1 standard error, N=979.}
\label{fig:change}
\end{figure}

In line with our prediction, we find that energies of belief networks estimated before the interventions can predict which individuals are most likely to change their beliefs during the interventions. In the study, participants were divided into five experimental groups for the GM food study and four experimental groups for the childhood vaccines study (and one control condition in each study, where participants did not receive any intervention; we excluded this control condition from the current analysis), each of which received the same scientific message about safety with a different framing (for the full list of educational interventions, see Supplementary Table \ref{tab:sq2} in Supplementary Materials). We calculated correlations between estimated energies and belief change separately for GM food and childhood vaccines for each experimental intervention group in order to account for potential differences in responses to the framing. Meta-analyses combining all interventions showed weak positive correlations between energies and absolute change for both GM food (see Figure \ref{fig:change}a), r = .12, p = .01, and childhood vaccines (see Figure \ref{fig:change}b), r = .16, p = .004 (see Supplementary Table \ref{tab:S2} for correlations per experimental group). Individuals with higher belief network energies were more likely to change their beliefs compared to individuals with low belief network energies. In other words, individuals who had less consistent beliefs before the intervention were more likely to change their beliefs after the intervention. In contrast, individuals who had more consistent beliefs were unlikely to change their beliefs.

As predicted by our cognitive network model, individuals who changed their beliefs were more likely to move to lower energy states. We tested whether individuals’ belief network energies were lower after the intervention compared to before the intervention. Meta-analyses combining all intervention groups showed that indeed energies were lower after the interventions for both GM food (see Figure \ref{fig:change}c), d = .34, p < .001, and childhood vaccines (see Figure \ref{fig:change}d), d = .33, p < .001 (see Supplementary Table \ref{tab:S3} for differences per experimental group). These effects were weak to moderate. Taken together, these analyses indicate that interventions aiming at changing individuals’ beliefs lead to reconfiguration of beliefs allowing individuals to move to lower energy states and more consistent belief networks. These re-configurations, however, can be either in line with the intervention’s aim or a backlash. In additional analyses, we found that reduction of energy is associated with absolute belief change even when controlling for other individual-level characteristics across educational interventions (Supplementary Materials Figure \ref{fig:predictmodl}). 

\section*{Discussion}
In this paper, we showed that formalising dissonance as energy of a belief network can be useful to predict and understand mechanisms leading to belief change. We proposed a cognitive network model which combines both social and moral beliefs and tested it using a longitudinal survey. This model enabled us to precisely estimate important predictors of belief change, such as the relationship between beliefs (couplings), their overall dissonance (energy), and the randomness of beliefs (temperature). Expanding on the AE framework \cite{Dalege2018}, we estimated cognitive networks combining social and moral beliefs. Using a longitudinal nationally representative study, we found full or partial support for our three predictions derived from the model. First, in line with our prediction, we found that attention to beliefs during the study was associated with a decrease in estimated temperature of the network. Second, we found partial support for our prediction that individual estimated energies should positively relate to self-reported feelings of dissonance. This relation was found for participants with positive beliefs towards GM food or childhood vaccines. Finally, we presented evidence in support of our third prediction: Individual-level energies were related to belief change after an educational intervention and this belief change lead to lower energy states. While our cognitive network model is only an analogy for actual cognitive processes, these findings show its usefulness in estimating and disentangling key psychological factors influencing belief change. 

We have two main contributions. Our first main contribution is to combine social and moral beliefs into a single cognitive network model built through a statistical physics framework. This model extends our recent framework for unifying moral and social beliefs \cite{Galesic2020} by also taking the network structure of all beliefs into account. Additionally, this model draws on previous research combining moral and social beliefs \cite{Galesic2019} and network models on relationships between beliefs \cite{Schweighofer2020, Rodriguez2016}. Previous research on belief formation and change have stressed the importance of both these sets of factors as individuals make decisions. Due in part to lack of cross-disciplinary research, however, the combination of both sets in one framework remains rare. In this paper, we draw on social psychology and statistical physics to not only incorporate beliefs across these two domains, but include them as part of an interacting network. We hope this research encourages more studies of the interactions between social and moral belief networks as important determinants for belief change.

Our second main contribution is that our cognitive network model is able to empirically predict belief change by connecting physical parameters to actual psychological constructs. Many belief dynamic models have remained untested on empirical data. In addition to a formal model, we provide empirical predictions about belief change using data collected specifically to answer these questions. Using a model based in social psychology, we bridge the gap between belief dynamics models in statistical physics and empirical work on science communication. We develop clear psychological meanings for statistical physics parameters and test their empirical validity. Belief network energies provide a formalisation of dissonance and temperature provides a formalisation of attention directed at an issue. This enables us to illuminate some of the mechanisms behind belief change: Individuals are motivated to reduce dissonance between beliefs and reconfigure their beliefs to allow lower dissonance. Such reconfiguration can be, but is not necessarily, in line with the aim of the intervention. The direction in which individuals change their beliefs does not only depend on the intervention but also on the easiest way for individuals to reduce their dissonance. This finding also goes above and beyond the classic finding that inducing dissonance leads to belief change \cite{aronson1959effect, egan2007origins, harmon2002testing} by showing that providing individuals with new information interacts with dissonances in their belief network. Individuals with low dissonance are unlikely to change at all, while individuals with high dissonance can change in both directions.

There are some limitations to the study. First, we estimated temperature per time point for the whole group of participants because current network estimation methods are not able to estimate temperature separately for each individual. The group-level network temperature thus likely represents the average temperature of the group with individual variation possibly captured by variations in energy. A longer longitudinal study and more advanced methods would enable individual-level estimates for temperature. Second, we did not have an empirical measure of attention and so could only infer that our estimated measure of temperature was related to attention through other proxies. However, temperature could reflect many psychological processes leading to the likelihood of belief networks moving to more consistent states or not. Third, as discussed above, our model predicted absolute belief change but not the direction of belief change, towards either acceptance or rejection of the safety of GM food and vaccines. Future research should expand on this model to provide ways to explain why some individuals accept or reject an experimental intervention and if individuals are in fact choosing the ``easiest'' path to a more consistent belief network. Finally, we focused on cognitive beliefs of one individual at a time, however, individuals are connected within larger social networks which influence the dynamics of belief change over a large population. We hope that subsequent research will continue to bridge social psychology and statistical physics to model and test belief change at the individual and societal level. 

This research has implications for science communication regarding issues critical to the health of many. We expect that scientific educational interventions that focus on reducing the belief network’s dissonance will be more effective in changing the minds of science sceptics. This applies specifically to the case of beliefs about GM food and vaccines but can be expanded to many other scientific issues, such as climate change. This study shows that given enough attention to the issue, individuals do change their mind if this enables less dissonance between all their beliefs within their cognitive network. Science communication should take into account how different moral and social beliefs are connected to each other to draft educational interventions that could lower the dissonance of the belief network in a way that leads to more acceptance of scientific facts. Further investigations to translate our findings to science communication might help combating erroneous and socially-detrimental beliefs.

\section*{Materials and Methods}
\subsection*{Experimental design}
We conducted a longitudinal study with an experimental component over three waves on a probabilistic national sample in the United States. To select participants for the study, we screened N=2,482 participants for their beliefs about the safety of GM food and childhood vaccines. We selected N=1,832  individuals who were somewhat hesitant about the safety of GM food or vaccine for the main experimental study. In other words, we only included individuals who selected a number between 1 and 6 (included) for the screener question “Do you think it is unsafe or safe to eat GM food?” or “Do you think childhood vaccines are unsafe or safe for healthy children?” with the options from 1-completely unsafe to 7-completely safe. 

Of the 1,832 selected participants, 979 completed the three waves with no missing values on any relevant questions and we only included these participants who had no missing values in our analyses. The first wave, on average 90 days after the screener, questioned participants about their beliefs about the safety of GM food and childhood vaccines as well as about related moral concerns and perceived beliefs of social contacts and sources. These questions were then administered again in wave 2, on average 20 days later, both before and after an experimental manipulation, and again in wave 3, on average 10 days later. 

To measure individuals’ moral and social beliefs about GM food and childhood vaccines, we included questions about related moral beliefs \cite{Graham2011}, and participants' perception of the beliefs of relevant social groups \cite{Galesic2018b}. Haidt \cite{Haidt2010} identifies six moral foundations relevant for different groups of U.S. Americans: Care, Fairness, Loyalty, Authority, Purity, and Liberty. We developed two questions for each of the moral foundations. For the social network, we focused on perceived beliefs about the safety of GM food or vaccines from direct social contacts (family and close friends, online community) and relevant sources of information (medical doctors, scientists, governmental agencies, online influencers, journalists, and the US general public). Full list of questions focused on moral and social beliefs are in Supplementary Table \ref{tab:sq1}. 

We included other questions relevant for the model in each wave of the questionnaire. First, we developed three questions focused on felt dissonance. These questions asked if the participant felt at ease, unbothered, and comfortable (all also on a scale from 1 to 7 and recoded so that higher values indicate higher dissonance). We averaged these three items into an index of felt dissonance. Cronbach's alphas in the different waves were high for both GM foods and childhood vaccines (GM foods wave 1: .93, wave 2: .93, wave 3: .94; Childhood vaccines wave 1: .92, wave 2: .93, wave 3: .95), indicating high reliability. 

In the second survey wave, participants were randomised into different experimental groups that received scientific facts about GM food and vaccines combined with messages targeting different social and moral considerations. The Supplementary Materials include the experimental conditions for participants selected for the GM food study (N=549) and the childhood vaccines study (N=430) (see Supplementary Table \ref{tab:sq2}). Each message within the GM and vaccines surveys had similar levels of readability and word count.

\subsection*{Network estimation and calculation of energies}
We estimated belief networks including moral and social beliefs for GM food and childhood vaccines, respectively. Before estimating the networks we regressed each belief on person and time to partial out these effects. We then used the residuals of these regression analyses to estimate the networks. We implemented our theoretical model using the Gaussian Graphical Model (GGM), which is the most common approach to estimate networks from continuous data. Edges in a network represent partial correlations between two nodes while controlling for all other nodes. Modelling the variance-covariance matrix $\Sigma$ can be done in the following way \cite{Epskamp2017}: 
\begin{equation} \label{eq:3}
\Sigma = \Delta(I-\Omega)^{-1}\Delta
\end{equation}
where $\Omega$ represents the partial correlations between nodes and measures the couplings $\omega$ of our cognitive network model. $\Delta$ represents a diagonal scaling matrix with square roots of the diagonal precision matrix scaling the partial correlations on the diagonal and 0s on the off-diagonal. These scaling values measure the temperature $\frac{1}{\beta}$ of our model. The difference between these scaling values and temperature is that there is one scaling value for each belief, while there is a single value for temperature in our model. The reason to have a separate scaling value for each belief is that scaling a GGM by a single value often results in a variance-covariance matrix that is not positive definite. As is the case for temperature, lower scaling values result in higher correlations between beliefs, because the model-implied correlations result from dividing the partial correlation between two given beliefs by the product of their scaling values. The average of these scaling values can therefore be regarded as a measure of temperature.    

We fitted networks separately for GM foods and childhood vaccines across the different time points and increasingly constrained the parameters of the specifications of our model in several steps and assessed the fit of these different specifications based on the Bayesian Information Criterion (BIC). These specifications were estimated using the R-package psychonetrics \cite{Epskamp2020}. We compared the fit of eight specifications of our model with increasing constraints of the estimated networks. We first let all parameters vary freely across time points and subsequently constrained the following parameters to be equal across time points: partial correlations between nodes ($\Omega$), intercepts (mean values) of the nodes, and scaling values ($\Delta$, as a proxy of temperature). We include constraints in the intercepts, because this allows us to use an approach similar to testing measurement invariance and makes variations in the scaling values identifiable. We tested each constraint using either a dense (all nodes being connected) or sparse network (some couplings set to 0). We determined which couplings were set to 0 using a prune-step-up procedure, which sets a given coupling to 0 and tests whether this results in better or worse model fit. We then selected the best fitting specification of the model.

For both the GM food and the childhood vaccines networks, the best fitting specification of our model was a sparse model (i.e., some partial correlations between beliefs were set to be 0) with equal partial correlations across time points (i.e., partial correlations between all beliefs were set to the exact same values at every time point)  and intercepts (mean values) but unconstrained temperature across time points (see Supplementary Table \ref{tab:S1} for fit measures of the different specifications of the model), implying that the network structure remained constant over time, while temperature varied over time. 

To calculate belief network energies per person at each time point, we used estimated partial correlations of the belief network. We multiplied the partial correlation between any given two beliefs with the recorded responses each individual had on these beliefs. The belief network energy is then the sum of these pairwise energy scores multiplied by -1.

\subsection*{Test of assumptions and validation}
To test the appropriateness of our assumption that Gaussian distributions were fitting our data, we investigated whether the multivariate distributions of the measured beliefs confirmed to a normal distribution. As can be seen in Supplementary Figure \ref{fig:normal} this assumption was met. 
To test the appropriateness of our assumption that couplings estimated at the group level were representative of couplings at the individual level, we first investigated the relationship between individual and group variances. The results of this analysis are shown in Supplementary Figure \ref{fig:indivpop}. We found that questions with high individual-level variance over time tended to also have higher group-variation at one time point. Second, we compared correlations between beliefs estimated over all time points without taking into account the multi-level nature of our data with correlations controlling for the multi-level nature of our data (with time points nested within individuals). As can be seen in Supplementary Figure \ref{fig:heatmaps} the relative size of correlations between beliefs was remarkably similar between the different forms of estimations, indicating that group-level correlation were similar to individual-level correlations. This was also confirmed by almost perfect correlations between the different estimation methods (GM foods: r (188) = .98, p < .001; childhood vaccines: .97 (188), p < .001). Only the absolute size of the correlation coefficients was affected by the different estimation methods. Correlation coefficients were considerably higher when the multi-level nature of the data was not taken into account (GM food: mean r = .39; childhood vaccines: mean r = .41) than when it was taken into account (GM food: mean r = .24; childhood vaccines: mean r = .28). Taken together, we conclude that it is likely that couplings estimated at the group level were representative of couplings at the individual level. 

In order to further test the validity of the model, we compared estimated and self-reported centrality of beliefs. Investigating the relationship between estimated and self-reported centrality allowed us to test whether the estimated network structure is in line with the subjective perception of the participants. For this analysis, we made use of additional questions in our data. Participants rated how important their different beliefs are for their belief about the safety of GM food or childhood vaccines. Participants also rated to what extent they believed that GM food or childhood vaccines are safe. For this analysis, we re-estimated the belief network including the safety beliefs. Results of this analysis are shown in Supplementary Figure \ref{fig:centralityImportance}. We found that estimated measures of centrality in relation to the safety of GM food and childhood vaccines correlated with self-reported importance of moral and social beliefs. 

\subsection*{Statistical analysis}
\emph{Correlation between energies and absolute belief change}. To test whether belief network energies predict belief change, we first correlated belief network energies and absolute belief change separately for each intervention and for each topic (these correlations and their associated significance levels can be found in Supplementary Table \ref{tab:S2}). We then transformed these Pearson's correlation into Fisher's z-scores and entered the scores into a random-effects meta-analyses separately for GM food and childhood vaccines. Finally, we re-transformed the Fisher's z-scores back to correlation coefficients for ease of interpretation. 

\emph{Differences between energies before and after the interventions}. To test whether belief network energies decrease after the interventions, we first calculated the mean differences between energies before and after each intervention separately for each topic (these mean differences and their associated significance levels can be found in Supplementary Table \ref{tab:S3}). We then transformed these scores into standardised mean change scores and entered the scores into a random-effects meta-analyses separately for GM food and childhood vaccines. Finally, we re-transformed the standardised mean change scores back to raw differences for ease of interpretation.   

\pagebreak

\bibliography{references.bib}   

\pagebreak

\section*{Acknowledgements}
The authors thank Mirta Galesic, Artemy Kolchinsky, Jens Lange, and Henrik Olsson for comments and discussions on earlier drafts. 

\subsection*{Funding}
National Science Foundation grant BCS-1918490 (JD) \\
EU Horizon 2020 Marie Curie Global Fellowship No. 889682 (JD) \\
National Science Foundation grant DRMS-1949432 (TvdD) \\
National Institute of Food and Agriculture 2018-67023-27677 (TvdD) 

\subsection*{Author contribution} 
Conceptualization: JD, TvdD \\
Methodology: JD, TvdD \\
Investigation: JD, TvdD \\
Visualization: JD, TvdD \\ 
Supervision: JD, TvdD \\ 
Writing--all: JD, TvdD 

\subsection*{Competing interests}
There are no competing interests to report. 

\subsection*{Data and materials availability} 
All data is available after registration on RAND American Life Panel's public depository: \href{https://alpdata.rand.org/index.php?page=data}{https://alpdata.rand.org}. 
Study numbers: 524-526. All code is available upon request. 

\pagebreak

\section*{Supplementary Materials}

\setcounter{table}{0}
\setcounter{figure}{0}
\setcounter{equation}{0}

\renewcommand{\figurename}{Supplementary Figure}
\renewcommand{\tablename}{Supplementary Table}

\begin{longtable}{|p{0.5\linewidth}|p{0.5\linewidth}|}
\caption{Questions about moral and social beliefs related to scientific issues (GM food or childhood vaccines).}
\label{tab:sq1}%
\endfirsthead
\endhead
    \toprule
    \textbf{Moral} & \textbf{Social} \\
    \midrule
    Instructions: For each of the following considerations, select a number between 1 and 7 that most closely represents your opinion. & Instructions: As far as you know, what is the percentage of people within the following groups who believe \textit{scientific issue} is safe to eat for healthy children? Please give your best estimate. \\
    \midrule
    \textit{scientific issue} is not part of our tradition, or that it is part of our tradition? & \% of medical doctors believe \textit{scientific issue} is safe. \\
    \midrule
    Producing \textit{scientific issue} is harmful or beneficial to the environment? & \% of representatives of governmental agencies believe \textit{scientific issue} is safe. \\
    \midrule
    \textit{scientific issue} negatively or positively affects your family? & \% of my family and close friends believe \textit{scientific issue} is safe. \\
    \midrule
    \textit{scientific issue} is not approved by all the appropriate agencies, or that it is approved by all the appropriate agencies? & \% of the US general public believe \textit{scientific issue} is safe. \\
    \midrule
    Some important information about \textit{scientific issue} is not shared with the public, or that all important information is shared? & \% of my online community believe \textit{scientific issue} is safe. \\
    \midrule
    Big biotechnology/pharmaceutical companies benefit more from \textit{scientific issue} than farmers/patients, or that both big biotechnology/pharmaceutical companies and farmers/patients benefit from \textit{scientific issue}? & \% of online experts and influencers I follow believe \textit{scientific issue} is safe. \\
    \midrule
    Large-scale farmers/medical doctors benefit more from \textit{scientific issue} than small-scale farmers/patients, or that both large-scale/medical doctors and small-scale farmers/patients benefit from \textit{scientific issue}? & \% of US scientists believe \textit{scientific issue} is safe. \\
    \midrule
    \textit{scientific issue} is unnatural or natural? & \% of journalists at your favorite news outlet believe \textit{scientific issue} is safe. \\
    \midrule
    You are forced to eat/use \textit{scientific issue} without your consent, or that you are free to choose whether to eat/use \textit{scientific issue}? &  \\
    \midrule
    God disapproves or approves of \textit{scientific issue}? &  \\
    \midrule
    \textit{scientific issue} is harmful or beneficial to children? &  \\
    \midrule
    \textit{scientific issue} negatively or positively affects our country? &  \\
    \bottomrule
\end{longtable}%

\pagebreak

\begin{longtable}{|p{0.5\linewidth}|p{0.5\linewidth}|}
\caption{Experimental groups and educational interventions.}
\label{tab:sq2}%
\endfirsthead
\endhead
    \toprule
    \textbf{GM experiment} & \textbf{Vaccines experiment} \\
    \midrule
    Nothing N=102 & Nothing N=89 \\
    \midrule
    Simple message N=90 & Simple message N=84 \\
    There is no evidence that GM food currently on the market is harmful when consumed by people or farm animals. This information comes from a summary of recent studies compiled by the National Academies of Science, Engineering, and Medicine. & There is no evidence that currently recommended childhood vaccines cause long-term harm to healthy children. This information comes from a summary of recent studies compiled by the National Academies of Science, Engineering, and Medicine. \\
    These studies investigated evidence about the effects of GM food when consumed by people or farm animals. This summary comes in the form of a written report. & These studies investigated evidence about the effects of childhood vaccines on children’s health. This summary comes in the form of a written report. \\
    \midrule
    Scientists N=86 & Scientists N=82 \\
    There is no evidence that GM food currently on the market is harmful when consumed by people or farm animals. This information comes from a summary of recent studies compiled by the National Academies of Science, Engineering, and Medicine. & There is no evidence that currently recommended childhood vaccines cause long-term harm to healthy children. This information comes from a summary of recent studies compiled by the National Academies of Science, Engineering, and Medicine. \\
    A survey of the largest scientific society in the country revealed that 88 \% of US scientists believe that it is safe to eat GM food. & A survey of the largest scientific society in the country revealed that 86 \% of US scientists believe that all children should be required to be vaccinated. \\
    \midrule
    Tradition (authority) N=90 & Tradition (authority) NA \\
    There is no evidence that GM food currently on the market is harmful when consumed by people or farm animals. This information comes from a summary of recent studies compiled by the National Academies of Science, Engineering, and Medicine. & \multicolumn{1}{r|}{} \\
     The production of GM crops imitates traditional farming techniques. These studies explain how all food crops have been modified from their wild relatives by farmers for thousands of years. & \multicolumn{1}{r|}{} \\
    \midrule
    Farmers (fairness) N=90 & Big corporations (fairness) N=88 \\
    There is no evidence that GM food currently on the market is harmful when consumed by people or farm animals. This information comes from a summary of recent studies compiled by the National Academies of Science, Engineering, and Medicine. & There is no evidence that currently recommended childhood vaccines cause long-term harm to healthy children. This information comes from a summary of recent studies compiled by the National Academies of Science, Engineering, and Medicine. \\
    Both big biotechnology companies and farmers benefit from GM crops. These studies show how adopting GM crops increases yield and reduces costs for both large-scale and small-scale farmers. & Big pharmaceutical companies do not benefit from vaccines at the expense of patients. These studies explain how companies would profit more from people getting sick than from selling vaccines. \\
    \midrule
    Information (freedom) N=91 & Information (freedom) N=87 \\
    There is no evidence that GM food currently on the market is harmful when consumed by people or farm animals. This information comes from a summary of recent studies compiled by the National Academies of Science, Engineering, and Medicine. & There is no evidence that currently recommended childhood vaccines cause long-term harm to healthy children. This information comes from a summary of recent studies compiled by the National Academies of Science, Engineering, and Medicine. \\
    This information is shared with the public. This organization maintains a website which includes a list of reports on research conducted about GM food (https://www.nap.edu/topic/298/ \newline
    agriculture/crop-and-plant-production). & This information is shared with the public. This organization maintains a website which includes a list of reports on research conducted about childhood vaccines. (https://www.nap.edu/collection/55/
    \newline
    vaccines) \\
    \bottomrule
\end{longtable}%

\pagebreak

\begin{table}
\centering
\caption{Fit measures of network model specifications.}
\label{tab:S1}
\begin{tabular}{lllll}\hline
Model  & DF & BIC & DF & BIC \\
& (GM food) & (GM food) & (Childhood vaccines) & (Childhood vaccines)\\
\hline
All parameters free & 0 & 11162.88 & 0 & 4236.69\\
(dense) & & & \\
All parameters free & 557 & 7546.51 & 552 & 768.09\\
(sparse) & & & \\
Equal networks & 570 & 7557.98 & 570 & 863.80\\
(dense) & & & \\
Equal networks & 679 & 6862.13 & 680 & 164.47\\
(sparse) & & & \\
Equal networks and & 630 & 7096.31 & 630 & 406.94\\
thresholds (dense) & & & \\
Equal networks and & 739 & 6400.47 & 740 & -282.53\\
thresholds (sparse) & & & \\
All parameters equal & 690 & 8275.81 & 690 & 1507.71\\
(dense) & & &\\
All parameters equal & 798 & 7567.16 & 801 & 167.19 \\
(sparse) & & &\\
\hline
\end{tabular}
\begin{tablenotes}
     \item[1] DF: Degrees of Freedom due to constraints of the model specifications.
     \item[2] BIC: Bayesian Information Criterion.
        \end{tablenotes}
\end{table}

\begin{table}
\centering
\caption{Correlations between energies and absolute belief change}
\label{tab:S2}
\begin{tabular}{ll}\hline
Intervention & Correlation\\
\hline
\underline{Gm food} & \\
Information & r=0.12, t(87)=1.11, p=0.27\\
Farmers & r=0.00 t(88)=-0.03, p=0.98\\
Scientists & r=0.17, t(83)=0.53, p=0.12\\
Tradition & r=0.22, t(88)=2.09, p=0.04\\
Simple & r=0.15, t(88)=1.42, p=0.15\\
\underline{Childhood vaccines} & \\
Information & r=0.06, t(84)=0.56, p=0.57\\
Scientists & r=0.12, t(80)=1.04, p=0.30\\
Simple & r=0.15, t(82)=1.41, p=0.16\\
Big corporations & r=0.30, t(86)=2.89, p=0.004 \\
\end{tabular}
\end{table}

\begin{table}
\centering
\caption{Differences between energies before and after the interventions}
\label{tab:S3}
\begin{tabular}{ll}\hline
Intervention & Mean difference\\
\hline
\underline{Gm food} & \\
Information &  MD = 0.000, t(88)=1.57, p=0.12\\
Farmers &  MD = 0.001, t(89)=3.39, p=0.001\\
Scientists &  MD = 0.001, t(84)=3.33, p=0.01\\
Tradition &   MD = 0.002, t(89)=3.26, p=0.002\\
Simple &   MD = 0.002, t(89)=4.61, p<0.001\\
\underline{Childhood vaccines} & \\
Information & MD = 0.001, t(85)=3.85, p=0.005\\
Scientists & MD = 0.001, t(81)=2.67, p=0.001\\
Simple & MD = 0.002, t(83)=4.53, p<0.001\\
Big corporations & MD = 0.001, t(87)=2.51, p=0.01\\
\end{tabular}
\end{table}

\begin{figure}
    \centering
\includegraphics[width=\textwidth]{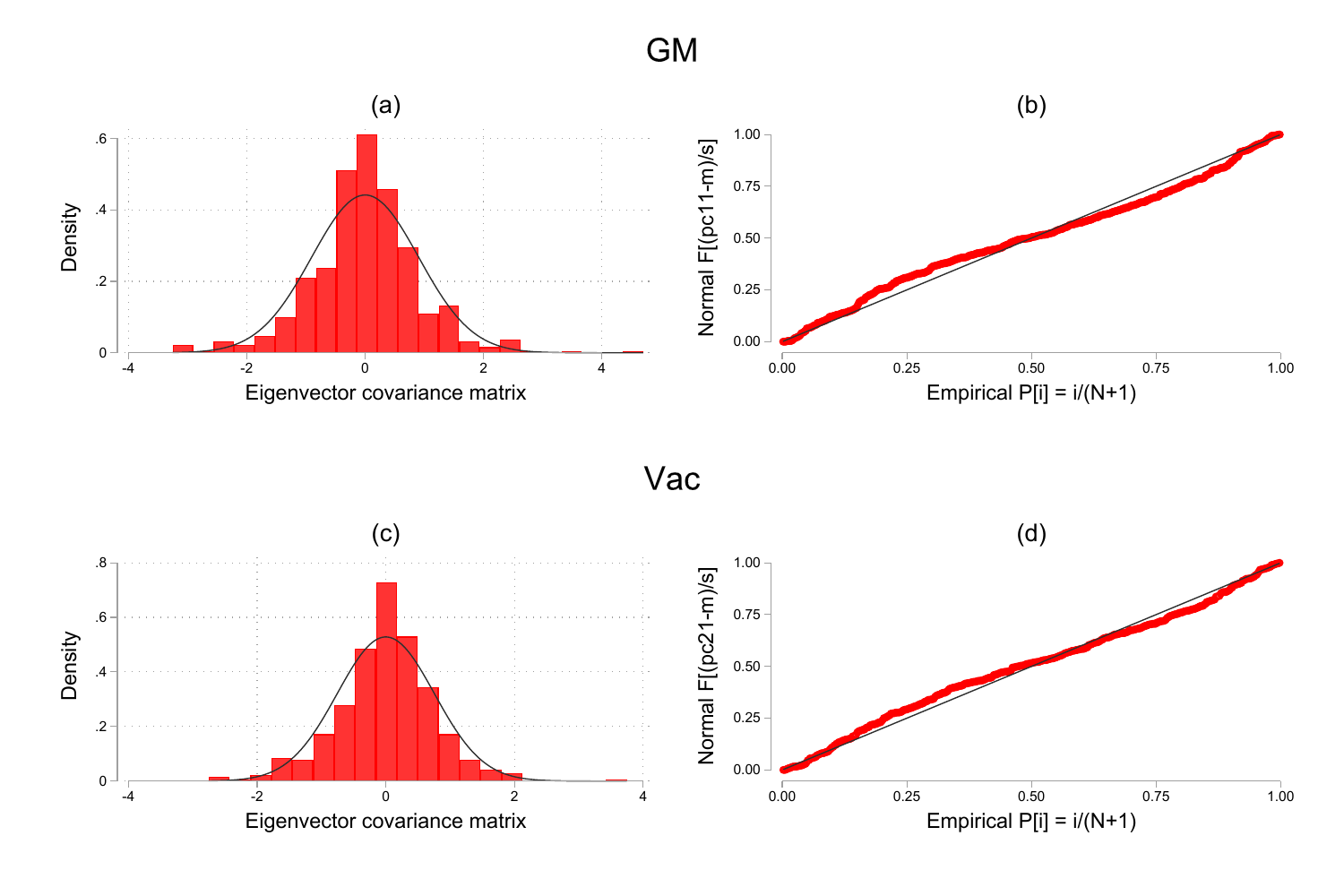}
\caption{Distribution of covariance matrix eigenvector (principal component of moral and social beliefs across time points) compared with the  normal distribution (a and c) and comparison of eigenvector with normal probability plot (b and d), for GM (a and b) and vaccines (c and d). In red are empirical data, N=979.}
\label{fig:normal}
\end{figure}

\begin{figure}
    \centering
\includegraphics[width=\textwidth]{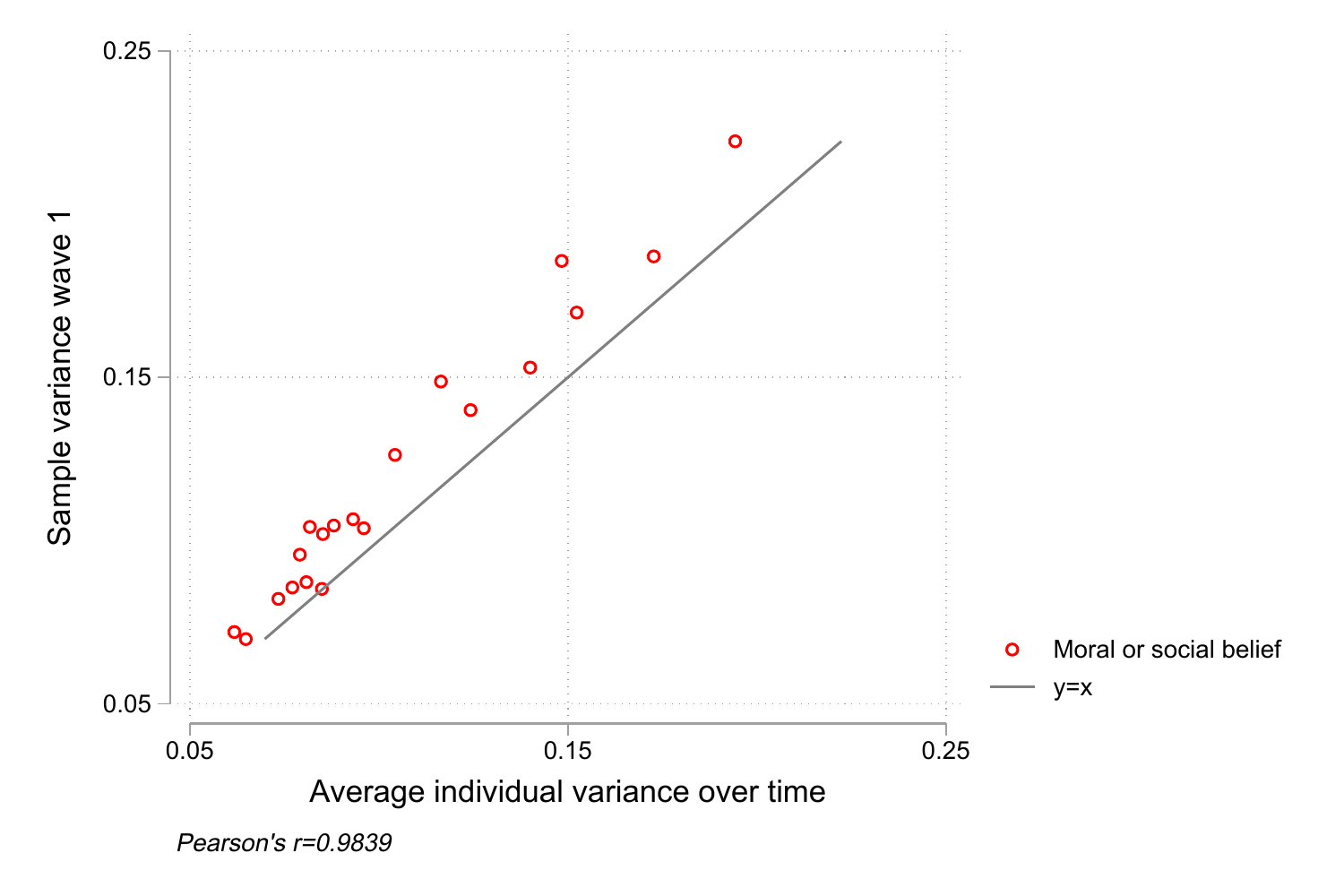}
\caption{Relationship between individual-level (over time) and population-level (across individuals at one time point) variances for each moral and social belief. Each dot represent one moral or social belief.}
\label{fig:indivpop}
\end{figure}

\begin{figure}
    \centering
\includegraphics[width=\textwidth]{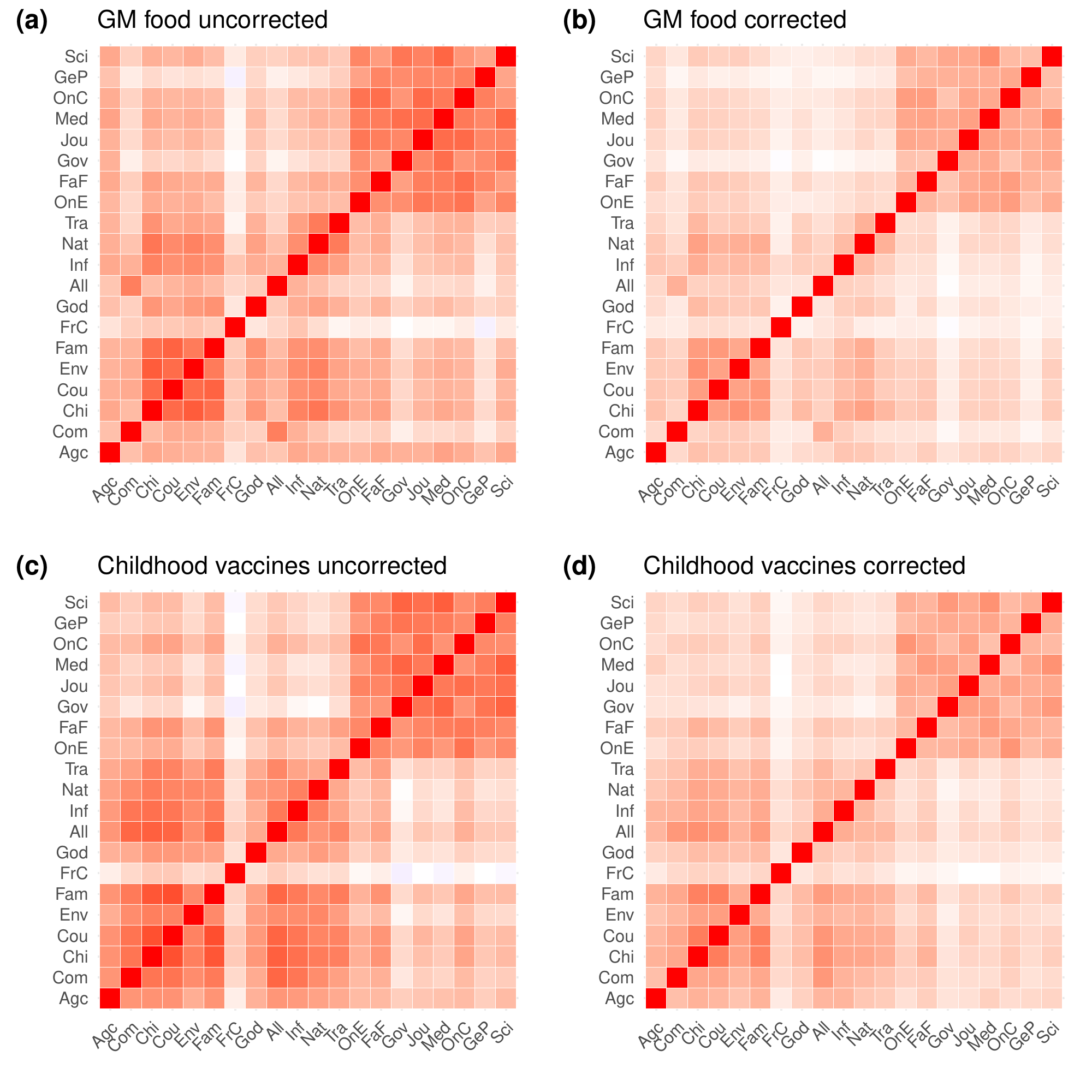}
\caption{Heatmaps of correlations uncorrected for multi-level nature of the data (a, c) and corrected of the multi-level nature of the data (b, d) between beliefs for GM food (a, b) and childhood vaccines (c, d). Higher density of redness represents stronger correlation coefficients.}
\label{fig:heatmaps}
\end{figure}

\begin{figure}
    \centering
\includegraphics[width=\textwidth]{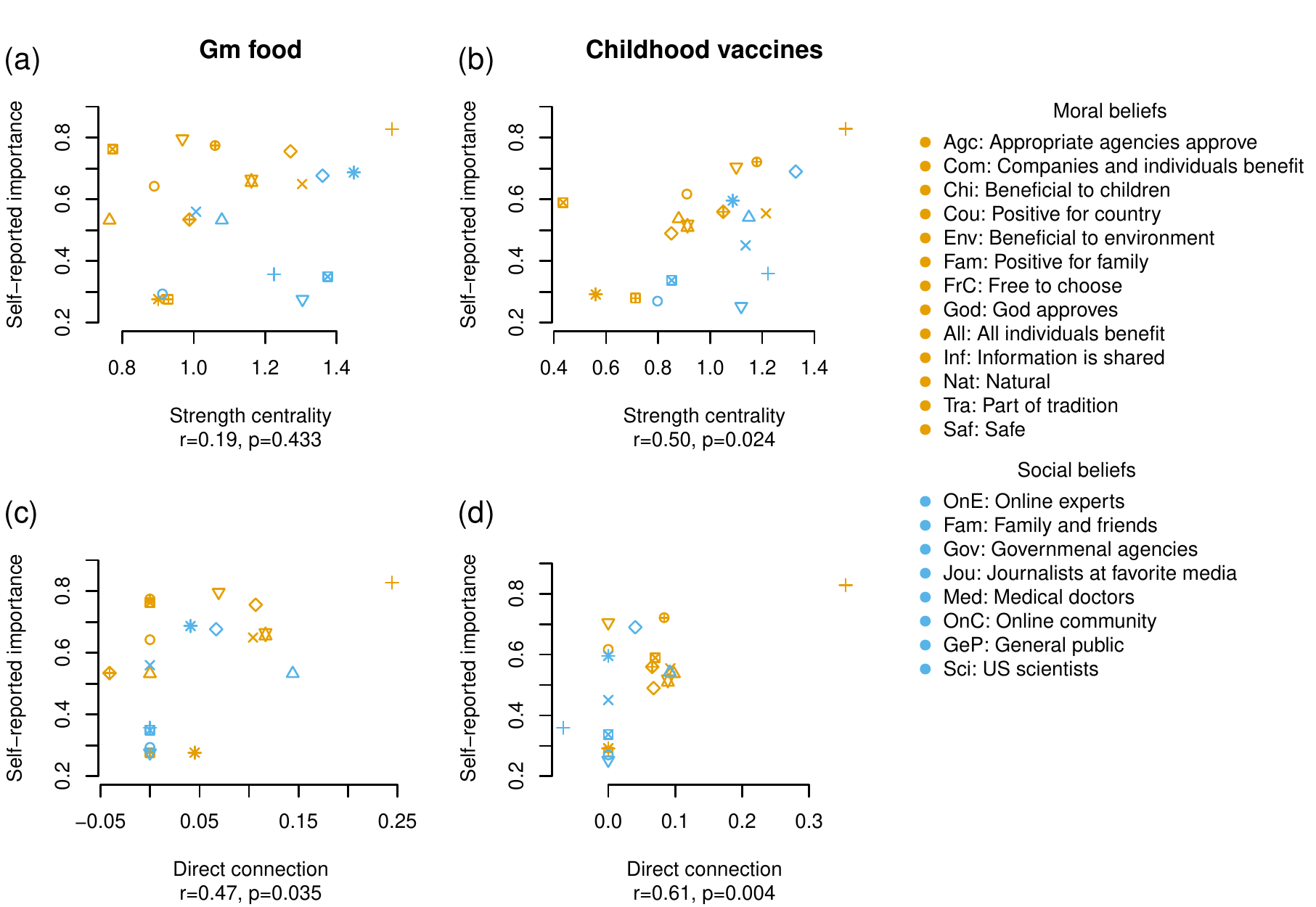}
\caption{Relationships between centrality of beliefs in the networks and self-reported importance of beliefs for the safety belief. (a) shows the relationship between strength centrality and importance for GM food. (b) shows this relationship for childhood vaccines. (c) shows the relationship between the magnitude of the direct connection to the safety belief and importance for GM food. (d) shows this relationship for childhood vaccines.}
\label{fig:centralityImportance}
\end{figure}

\begin{figure}
    \centering
\includegraphics[width=\textwidth]{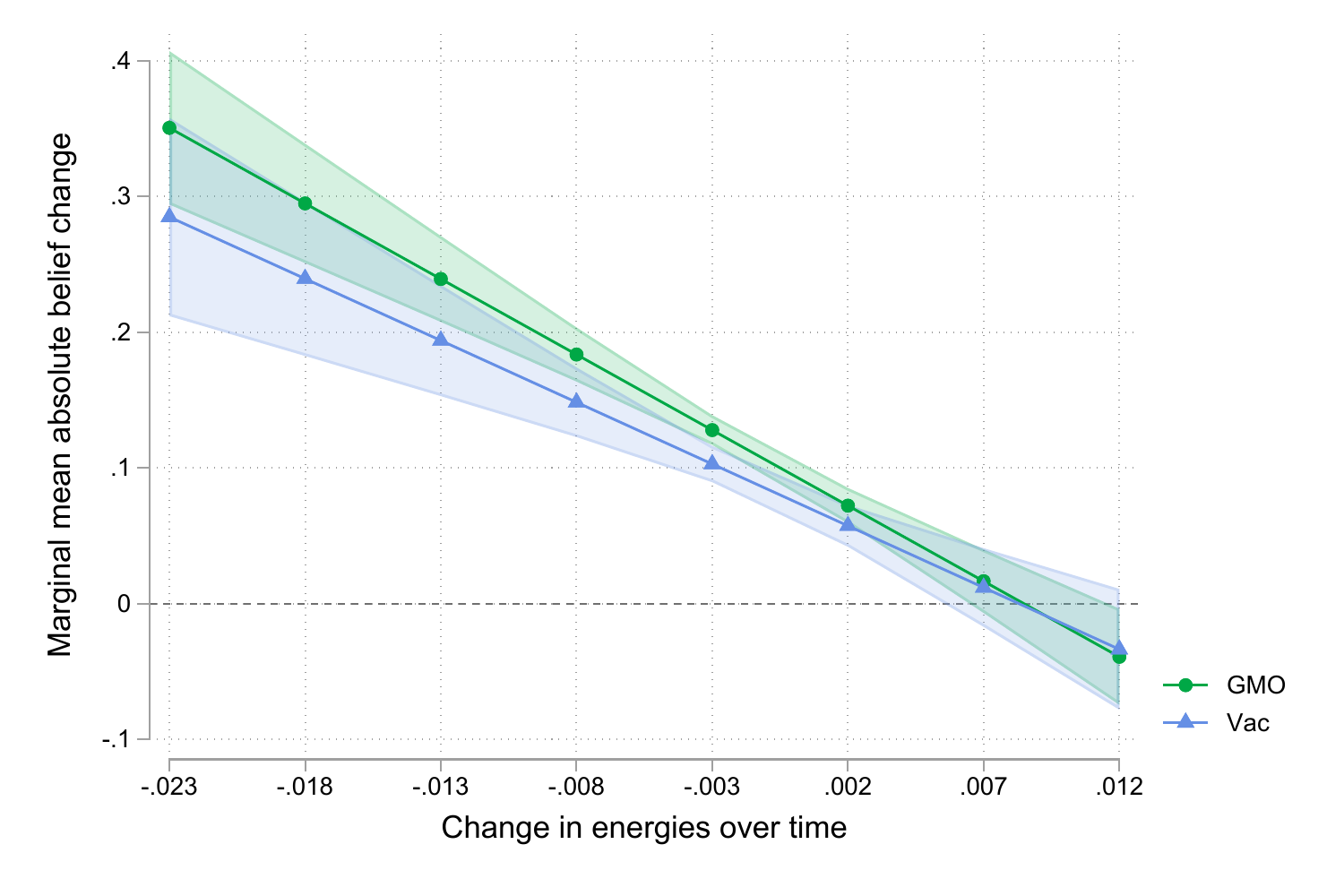}
\caption{Marginal mean absolute belief change given energy difference before and after the intervention. Predicted belief change estimated with experimental group, original beliefs, gender, education, family size and political views held at their sample values. Only including participants who received an educational intervention, N=788}
\label{fig:predictmodl}
\end{figure}

\end{document}